\documentstyle[11pt,aaspp4]{article}  

\begin{document}

\title{Eclipsing Binaries in the OGLE Variable Star Catalogs. V. \\
Long-Period $\mathbf{\beta}$~Lyrae-type Systems 
in the Small Magellanic Cloud and the PLC-$\mathbf{\beta}$ Relation}

\author{\sc Slavek M. Rucinski\\
\rm Electronic-mail: {\it rucinski@astro.utoronto.ca\/}}
\affil{David Dunlap Observatory, University of Toronto \\
P.O.Box 360, Richmond Hill, Ontario, Canada L4C~4Y6}
\and
\author{\sc Carla Maceroni\\
\rm Electronic-mail: {\it maceroni@coma.mporzio.astro.it\/}}
\affil{Rome Observatory,
via Frascati 33, I-00040 Monteporzio C., Italy}

\begin{abstract}
Thirty eight long-period ($P > 10$ days),
apparently contact binary stars discovered by the OGLE-II project
in the SMC appear to be $\beta$~Lyrae-type systems with 
ellipsoidal variations of the cool components 
dominating over eclipse effects in the systemic 
light variations and in the total luminosity. 
A new period--luminosity--color
(PLC) relation has been established for these systems; we 
call it the PLC-$\beta$ relation, to distinguish it from 
the Cepheid relation. Two versions of the PLC-$\beta$ relation --
based on the $(B-V)_0$ or $(V-I)_0$ color indices -- have been 
calibrated for 33 systems with $(V-I)_0 > 0.25$ spanning
the orbital period range of 11 to 181 days. The
relations can provide maximum-light,
absolute-magnitude estimates accurate
to $\epsilon M_V \simeq 0.35$ mag.\ within the approximate
range $-3 < M_V < +1$. In terms of their number in the SMC,
the long-period $\beta$~Lyrae-type binaries 
are about 50 times less common than the Cepheids.
Nevertheless, their large luminosities coupled with
continuous light variations make these binaries 
very easy to spot in nearby galaxies, so that
the PLC-$\beta$ relation can offer an auxiliary and
entirely independent method
of distance determination to nearby stellar systems rich
in massive stars. The sample of the long-period
$\beta$~Lyrae systems in the SMC analyzed in this paper is currently 
the best defined and uniform known sequence of such binaries.
\end{abstract}

\section{INTRODUCTION}
\label{intro}

This is the fifth paper based on the OGLE data, but the first
one based on the data from the second phase of the OGLE 
project (OGLE-II, \cite{ogleII}) utilizing the
new 1.3 meter Polish telescope on Las Campanas. The data
are freely available over the computer 
networks\footnote{http://www.astrouw.edu.pl/$\sim$ogle/ or
http://www.astro.princeton.edu/$\sim$ogle/.}.

Analysis of large numbers of eclipsing binary stars 
discovered by the OGLE-I project in the direction of 
Baade's Window (BW) revealed that among almost one thousand 
binaries, about two thirds were contact systems 
of the W~UMa-type (\cite{ogle1}, \cite{ogle2} = Papers I and II). 
The direction of the BW is towards the Galactic Bulge,
yet all W~UMa-type systems discovered there appear to be
Disk population objects distributed over the
long line of sight to the Bulge (\cite{open}).
While the W~UMa-type binaries are very short-period systems, 
with orbital periods within $0.25 < P < 0.6$ days (this
class of binaries is normally 
assumed to have periods shorter than one day), use of a 
Fourier decomposition on all available OGLE-I light curves 
showed that main-sequence contact binaries 
continue to exist in the Galactic Disk population
to orbital periods of about $1.3 - 1.5$ 
days and then abruptly vanish, with very few contact 
systems having longer periods (\cite{ogle3} = Paper III,
see Figure~2 in this paper). Only four apparently 
contact systems were discovered 
with the orbital periods $P > 10$ days. 
Because of the long periods, these could not be main sequence 
stars such as those in the typical W~UMa-type systems, 
but giants. Since there exist no theoretical predictions 
nor any observational indications that contact systems 
can be formed with giant components, no firm 
explanation for the long-period systems in BW could 
be offered at that time of Paper~III. However, it was 
noticed that their light variations could be rather 
easily be explained by orbital synchronous rotation 
of a {\it single\/} distorted component filling 
its Roche lobe. Thus, these long-period binaries could 
consist of just one evolved component and some other
star which -- because of its small size and/or low
temperature -- would be invisible, yet would
exert a sufficiently strong
gravitational pull to distort the visible component.

In this paper we suggest that the ``long-period contact binaries''
are in fact representatives of the well known group 
of $\beta$~Lyrae-type binaries. The primary basis 
for the argument is the well-populated sequence of long-period (LP) 
systems which has been established among contact
binaries discovered by the OGLE-II project in the Small 
Magellanic Cloud (SMC)(\cite{uda98a}). We have found members
of this sequence by using 
the same Fourier light-curve shape filter as in the
previous papers of this series. Although the filter
is meant to select genuine contact binary systems,
it can also pass binaries which are not in contact, but
which show sufficiently strong ellipsoidal 
variations and shallow or absent eclipses. 

The hypothesis the LP systems   
are in fact binary systems with only one, Roche-lobe filling 
component dominating in the total light
leads to a prediction of a period--luminosity--color (PLC) 
relation for such systems. 
Such a relation is expected to be very similar to what
has been suggested for the W~UMa-type stars (\cite{cal1}),
but with only one component defining the global relationship.  
If the total mass and the mass-ratio are not 
correlated with the orbital period (however, this is an
assumption only), 
a PLC relation will simply result from the Kepler-law 
scaling of the total systemic dimensions
(and thus of the stellar radius) with the orbital period, 
$R \propto P^{2/3}$. Further conversion to a luminosity
dependence, $L \propto R^2 T^4_{eff}$, and then to an
absolute-magnitude dependence, will lead to 
$M_V \propto -10/3 \, \log P$.  
In this paper we show that a sample of 
$\beta$~Lyrae systems in the SMC, 
selected with the exclusion of blue 
systems ($(V-I)>0.25$), forms a surprisingly uniform and 
coherent group for which an absolute-magnitude 
relation -- expressed as $M_V=M_V(\log P, \, color)$, with the expected
$\log P$ dependence -- can indeed be established.

We define the sample in Section~\ref{sample} and then 
show the period--color relation (Section~\ref{pc}) as the best
tool to delineate the sequence of the long-period contact
systems. In Section~\ref{lc} we show that in fact these
are $\beta$~Lyrae-type binaries, not genuine contact systems.
The color-magnitude diagram (Section~\ref{cmd}) leads
directly to a period--luminosity--color which we express as a
absolute-magnitude calibration (Section~\ref{cal}).
We discuss the OGLE-II data for the systems discovered
in the Baade's Window sample in Section~\ref{bw} and give
estimates of the spatial frequency of the LP systems
in Section~\ref{freq}. 
Section~\ref{disc} summarizes the results of this paper.

\section{SELECTION OF THE SAMPLE}
\label{sample}

The first step in the selection of the sample was
finding the contact systems among all eclipsing binaries
found by \cite{uda98a} in the SMC. Then, following
the results of the next Section~\ref{pc}, which suggest
a change in the sample characteristics at periods of
about 10 days, the long-period representatives of the sequence
were further selected. The $a_2 - a_4$ Fourier light-curve 
shape-filter has been applied to
all eclipsing binaries listed in \cite{uda98a}. 
The operation of the filter is shown in Figure~\ref{fig1}
while the details of the filter have 
been described before (Papers I, II, III and in \cite{mr99} 
= Paper IV). The filter requires 
the $\cos (4 \phi)$ term to be, in its absolute 
value, relatively small in relation to the $\cos (2 \phi)$
term ($\phi$ is the orbital phase). 
A line on the $a_2 - a_4$ relation, 
pre-computed for a case of the inner contact using
light curve synthesis models, 
defines a locus of systems with components just touching. 
Contact systems show small (in absolute value) $a_4$ 
coefficients when compared with $a_2$. In effect, 
the Fourier light-curve filter selects eclipsing binaries 
with dominance of geometric distortion effects 
(which define the overall amplitudes and control
the size of $a_2$) over eclipse effects 
(contributing primarily to the $a_4$ term). 

Normally, the Fourier filter works very well in correctly
identifying contact binaries. It also rejects
very effectively semi-detached binaries of the Algol type
which show deep eclipses. However, the filter
may give wrong results for a semi-detached binary with
very shallow or absent eclipses,
in the situation of a great disparity of sizes of the components,
with a large visible star showing a
considerable tidal distortion\footnote{The filter fails
also for systems with secondary minima displaced from
phase 0.5. Such systems had to be removed manually
form the sample.}. Then the $a_4$ coefficient
can also be small relative to $a_2$ and the 
binary may be erroneously classified as a contact system. 
This is exactly what we suspect
to be the case for the LP binaries described in this paper.

\placefigure{fig1}

As an aside, 
we note that the OGLE-II project eliminated small amplitude
systems from consideration at the early stage 
by exclusion of the so-called 
``ellipsoidal variables'' (\cite{uda98a}). 
This means that some small-amplitude contact, semi-detached
and perhaps even detached binaries 
have been excluded at the stage of formation of the 
sample. We consider this choice as somewhat unfortunate
because criteria for rejection were not clearly 
formulated in the discovery paper. Thus, the sample 
cannot be used for statistical studies 
involving light variation amplitudes or their derived
quantities (such as $a_2$ and $a_4$)
at the (poorly defined) low-amplitude ends of their distributions.

\section{PERIOD -- COLOR RELATION}
\label{pc}

The sequence of long-period contact systems is very well 
defined in the period-color (PC) relation which -- as was
many times stressed in this series -- plays a special
role for contact binary stars. The left panel of
Figure~\ref{fig2} shows the 
PC relation for the all SMC systems selected as contact
with the $a_2 - a_4$ light-curve shape filter. The 
sequence consists of 38 systems. It starts at 11 days
(in fact a couple of systems with periods of 8 -- 10 days
may also belong to the sequence)
and continues to 181 days. It has an inverted slope 
relative to the one normally observed for most common 
contact binaries of the W~UMa-type: Typical contact
binaries become hotter as their periods get longer, 
a trend which is explained by a progression upward 
along the main sequence. The inverted sense of the 
slope for the LP sequence suggests that they
become cooler with the increase of dimensions, a
property normally associated with giants. 

\placefigure{fig2} 

The SMC sample is not the first one to show the LP 
sequence: In addition to the indications from the 
OGLE-I Baade's Window (BW) sample 
that we mentioned in Section~\ref{intro},
the MACHO project has also detected 
several similar systems in LMC; the latter have been
analyzed in \cite{macho}. We 
show the LMC systems in the right panel of Figure~\ref{fig2}. 
Although the BW and the LMC samples give a 
qualitative support to what will be discussed in 
this paper, we cannot utilize the BW and LMC data 
together with the SMC data for the 
following reasons: (1)~It is has been found that the 
OGLE-I photometry for BW suffers from zero-point offsets as 
well as from scale errors resulting from the non-linearity
of the CCD detector. We discuss the details 
and present limited new OGLE-II data for three systems in
Section~\ref{bw}; (2)~The MACHO sample 
used unconventional spectral bandpasses 
which resulted in photometry only approximately 
transformable into the $(V,\, V-R)$ color system. 
Thus, neither the OGLE-I BW sample nor the MACHO 
sample can be compared with the new, high 
quality OGLE-II data for SMC. 

\section{LIGHT CURVES}
\label{lc}

The OGLE-II data for the SMC variable stars
consist of light curves in the $I$ band and of
the maximum light magnitudes and color indices $(B-V)$ and $(V-I)$.
The light curves are shown in Figures~\ref{fig3} and \ref{fig4}
while the photometric data are given in Table~\ref{tab1}. The first  
columns of the table contain the OGLE project data: First,
the identification numbers (the consecutive ones used
in this paper and then the OGLE-II numbers), followed
by the orbital periods in days and 
the three photometric quantities, the
$I_{max}$ magnitude and the color
indices $(B-V)$ and $(V-I)$. The OGLE-II identification number
is made of the SMC field number (integer part of the
number) and of the star number (six digits
after the period). Among the 38 systems found
in the SMC fields, three have additional identifications in
the neighboring fields. The collection of the
light curves in Figures~\ref{fig3} -- \ref{fig4} 
shows only the light-curve 
data based on the primary identifications. 
The light-curve Fourier decomposition coefficients $a_i$,
which are given in the last six columns of Table~\ref{tab1},
have been calculated for both identifications, but we
used the primary ones in selection of the contact binary
systems through the $a_2-a_4$ Fourier filter.
As in the previous papers of this series,
$a_i$ are the cosine coefficients while $b_1$ is the only
sine term representing light curve asymmetries.

\placefigure{fig3}

\placefigure{fig4}

The light curves shown in Figures~\ref{fig3} and \ref{fig4} 
are not of typical
contact systems with very similar depths of eclipses. All
show unequally deep eclipses which are characteristic 
for $\beta$~Lyrae (EB) light curves. 
The light curves are well defined for most systems. Twenty seven
systems with relatively large amplitudes of light 
variations have been selected to enter with full weight
in determinations of luminosty calibrations in Section~\ref{cal}.
Eleven systems with poorer light curves are marked by asterisks
in Figures~\ref{fig3} -- \ref{fig4} 
and in Table~\ref{tab1}; they are marked by
open circles in the remaining figures.

The similarity of all LP systems in terms of their 
light curves is confirmed by the $a_2-a_1$ plot in
Figure~\ref{fig5} where we see a relatively tight relation, 
almost a proportionality between the coefficients. For normal
W~UMa-type systems, the $a_1$ coefficients are confined
to a band $-0.02 < a_1 <0$, without any clear correlation
with $a_2$ (see \cite{ogle2}, Figure~5 there). Since the $a_i$ 
coefficients are defined in terms of light intensity
units, the proportionality of $a_1$ to $a_2$ implies
similar light curves in magnitudes, with
similar differences of the eclipse depths.
This is exactly the property that was noticed when an 
attempt was made to model the light curve of the LP 
system in the BW sample, BW0.036 (Paper III): 
For a system where the only visible component fills 
the Roche lobe, the relative depths of eclipses are 
expected to be very similar almost irrespective of 
the orbital inclination and only very slightly depend 
on the mass-ratio. 

\placefigure{fig5}

Although the light curves are similar, they also show
some small differences in that some show narrow secondary
eclipses while some show relatively more scatter at both maxima.
In one case, LP\#9, we see well defined eclipses indicating that
both components contribute to the systemic light; we show
later that this is exactly the system which deviates from the
PLC relation.
We have made no attempt to model the light curves of the 
LP systems. We feel that without spectroscopic 
mass ratios, this would be futile exercise. We note 
that color curves cannot be currently formed because
of much smaller numbers of the $B$ and $V$ observations
relative to the number of $I$ observations. However,
all the light curve features can be explained by
the same model which is applicable to the well-known
prototype of $\beta$~Lyrae.

At this point, we would like to make a clear 
distinction between the light-curve
shape designated by EB 
from the name of close-binary stars $\beta$~Lyrae.
From now on, $\beta$~Lyrae will
be used here as a binary-system prototype, not as a designator of 
of the EB light-curve shape. That is, when calling a star 
a $\beta$~Lyrae type, we will consider a binary with only one, 
giant-type component which dominates the total light, the other
component being a main sequence star or collapsed star
(perhaps surrounded by an accretion disk) 
with only a minor influence on the total systemic
brightness and its light curve. Such a model 
is now well established for the prototype, 
following the extensive research 
of $\beta$~Lyrae and of the related W~Serpentis-type binaries
(\cite{hua63}, \cite{woo65}, 
\cite{wil74}, \cite{pla85}, \cite{hub91}, 
\cite{har93}, \cite{deG94}) and of various X-ray sources with
giant-star mass donors. We stress that in calling a star
a $\beta$~Lyrae binary, we focus on two essential
properties: the semi-detached model and the giant characteristics
of the Roche-lobe filling component. Later on, to establish
an absolute-magnitude calibration, we add the third property:
the sufficiently yellow/red color which, in most
cases, indicates a negligible photometric 
importance of the other component.

The light curves of the LP $\beta$~Lyrae systems in the SMC sample
appear to be symmetric, with identical heights of the
light maxima (i.e. the so-called O'Connell effect is 
nonexistent for these systems). 
We also see no correlation between the sine 
coefficients, $b_1$, and the coefficients measuring 
the eclipse-depth differences, $a_1$. Such a correlation, 
manifesting itself as a preference for higher first maxima with the
increased eclipse-depth difference,
was observed for the short-period EB systems ($P < 1$ day),
appearing in the Baade's Window sample and 
analyzed in \cite{ogle2}. An explanation offered there 
was that these are semi-detached
binaries with matter being transferred from the more massive,
hotter to the less-massive, cooler components.  
Such systems are very different from the systems
discussed in this paper in that they do not have to
contain a giant, Roche-lobe filling component. In fact,
judging by short orbital periods, many of them consist of
main sequence stars first time evolving into the mass
exchange phase or, perhaps, experiencing the semi-detached
stage of the thermal-cycle relaxation oscillations
(\cite{luc76}, \cite{fla76}, \cite{RE77}, \cite{LW79}).

Finally, we note that
the light curve amplitudes of the LP systems, 
as measured by the $a_2$ Fourier coefficient, 
do not show any correlation with the
orbital period or the color index of the system. Note
also that the light curve shapes in Figures~\ref{fig3} -- \ref{fig4}
do not correlate in any obvious way with the periods
(the periods are given in parentheses above each panel
in these figures). In this respect, the sample appears to be uniform
in the sense of consisting of very similar systems.

\section{COLOR -- MAGNITUDE DIAGRAM}
\label{cmd}

The SMC data are an excellent material for forming a 
color-magnitude diagram (CMD) and for studying location 
of the LP systems in relation to the well recognized 
stages of stellar evolution. The distance modulus is 
relatively well known and will be soon known even better 
once the matter of systematic deviations between various
methods of distance determinations to Magellanic Clouds
is resolved (\cite{uda98}, \cite{uda98b}, \cite{gib99},
\cite{nel2K}, \cite{pop2K}). 
Also, the average reddening for the SMC is small. In contrast, 
the Baade's Window sample (Paper III) suffered from large and 
uncertain reddening corrections and from the 
(then unrecognized) photometric errors, more fully
described in Section~\ref{bw}. In addition, the distances of the BW
systems were of course different precluding formation 
of an independent color-magnitude diagram (other 
than the one involving an absolute-magnitude calibration). 
In principle, the 
MACHO sample for the LMC (\cite{macho}) could be used to 
form a CMD, but the data were in a non-standard 
photometric system and the reddening corrections 
(and their variations) are larger than typically for the SMC.

Figure~\ref{fig6} gives the observational CMD for the 
LP $\beta$~Lyrae
systems in the SMC, with the main stellar density enhancements 
(such as the main sequence or the red clump) marked by 
shading. The approximate discovery 
completeness limits of 90\% (the
slanting line starting at $I=17.5$ and $V-I=0$) and 
of close to 0\% (1.5 mag.\ below), following \cite{uda98a},
are shown by slanting, dotted lines. In transforming 
from the observed, maximum-light photometric quantities to
the de-reddened, absolute magnitudes, we used:
$(m-M)_0=18.8$, $E_{V-I}=0.12$ and $A_V=2.5 \, E_{V-I}$. 
The LP systems are much more luminous 
than the main-sequence stars. They are also systematically 
brighter than typical giants in the SMC; this may be partly a selection
effect as brightest stars are expected to have better photometry. 
Comparison with in evolutionary tracks in \cite{mas95}
indicates that the LP binaries considered in this paper
cover the area normally associated with evolved, 
moderately massive ($3-5 M_\odot$) stars within the range of
effective temperatures of $3.60 < \log T_{eff} < 4.00$. Later on,
we set a blue limit at $(V-I)_0=0.25$ so that the effective
temperature range is cut from above at $\log T_{eff} \simeq 3.9$.

The orbital periods of the LP $\beta$~Lyrae systems
are shown in Figure~\ref{fig6} by the size of the symbols. 
One immediately sees a period progression in that 
the systems with the longest periods appear in the
upper part of the CMD. The longest periods, $P > 100$ days,
appear exclusively among red giants above the Red Clump 
location, but a similar period progression is also visible in the 
upper-left, blue part of Figure~\ref{fig6} where 
five systems of similar color index have 
luminosities scaling with their orbital periods. This
type of the period dependence indicates that a
period -- luminosity -- color relation 
can be established for the LP systems.

\placefigure{fig6}

\section{PERIOD -- LUMINOSITY -- COLOR RELATION }
\label{cal}

\subsection{Calibration of the relation}

The previously established absolute magnitude 
calibrations for contact binary systems of the 
W~UMa-type are of limited value here, primarily 
because of the vastly different ranges of the 
orbital periods. While the W~UMa calibrations are 
applicable to the range 0.25 -- 0.6 days 
(\cite{rd97})\footnote{A blind application of the
$(V-I)$-based calibration of \cite{rd97}
to the SMC LP systems gives $M_V$ estimates about 3.5
magnitudes too bright.}, 
the calibration that we would like to establish 
will cover a large range of the orbital periods 
from 11 days to 181 days. 

We set aside and do not discuss the
contact systems with the intermediate periods, starting
around 0.6 -- 1.5 days at the short end and extending 
to 10 days. Such systems apparently do not exists in the 
solar neighborhood and are exceedingly rare in the BW sample 
(\cite{ogle3}), but -- possibly due to the discovery
selection effect preferring brightest objects -- are 
quite common in the present SMC and in the MACHO LMC samples 
(Figure~\ref{fig2}). We feel that the intermediate-period 
($1.5-10$ days) systems which are visible in the SMC and LMC samples
are a mix of genuine contact systems 
with the moderately-short representatives of the
$\beta$~Lyrae systems discussed in this
paper. They will be the subject of a separate investigation.

Assuming the same values for the
distance modulus and reddening as in the
previous section, the linear regression for the 
whole available data set of 38 systems (with 11 systems given
half weights) gives:
\begin{eqnarray}
M_V  & = & -3.90 \, \log P + 3.27 \, (V-I)_0 + 2.40, \; \; \sigma=0.50
  \label{eq1} \\
M_V  & = & -4.05 \, \log P + 3.01 \, (B-V)_0 + 3.46, \; \; \sigma=0.63
  \label{eq2}
\end{eqnarray}
The standard errors (per data point) $\sigma$ are 
large and one may question the validity of linear 
fits in so poorly constrained cases. The deviations from 
the linear regressions are shown in the upper two panels 
of Figure~\ref{fig7}. Note, that the bluest stars appear to 
deviate the most from the linear dependencies. 
If we exclude these systems setting the limit at 
$(V-I)0>0.25$, then the linear fits improve substantially
reaching the level of precision of about 0.35 mag.
This level of accuracy is similar to that for W~UMa-type binaries 
and for Cepheid variables. 
For the limited data set of 33 yellow-red systems, the linear weighted 
regression equations are:
\begin{eqnarray}
M_V & = & -3.43 \, \log P + 2.04 \, (V-I)_0 + 2.80, \; \; \sigma=0.34 
  \label{eq3}\\
M_V & = & -3.30 \, \log P + 1.49 \, (B-V)_0 + 3.38, \; \; \sigma=0.38
  \label{eq4}
\end{eqnarray}
The standard deviations are now quite respectably small 
so that the equations can be used for prediction of 
absolute magnitudes. The deviations from the linear
fits for this case are shown in Figure~\ref{fig8}.
Realistic uncertainties of the
coefficients for the sample of 33 yellow-red
systems have been established by application
of a bootstrap re-sampling process. 
The median values and the 1--$\sigma$ ranges give the following
relations:
\begin{eqnarray}
M_V & = & -3.41^{+0.28}_{-0.27} \, \log P + 
    2.00^{+0.29}_{-0.31} \, (V-I)_0 + 2.80^{+0.27}_{-0.27} 
  \label{eq5} \\
M_V & = & -3.29^{+0.34}_{-0.34} \, \log P +
    1.48^{+0.31}_{-0.34} \, (B-V)_0 +3.36^{+0.43}_{-0.39}
  \label{eq6}
\end{eqnarray}

\placefigure{fig7}

\placefigure{fig8}

The $\log P$ coefficients are very close what we suggested
would be the case on the basis of very simplified considerations
in Section~\ref{intro},
$M_V \propto -3.33 \, \log P$, which is highly gratifying. But we
achieved that by removing the five blue systems with 
$(V-I)0<0.25$. This removal is not entirely arbitrary;
we discuss it in the next subsection.
One can fix the period coefficient at the expected Kepler-law value of 
$-3.33$ and determine the other two regression coefficients with
higher much accuracy. The results for 33 yellow-red systems are:

\begin{eqnarray}
M_V + 3.33 \, \log P & = & 1.94 \, (V-I)_0 + 2.72, \; \sigma=0.34 
  \label{eq7} \\
M_V + 3.33 \, \log P & = & 1.51 \, (B-V)_0 + 3.40, \; \sigma=0.38 
  \label{eq8}
\end{eqnarray}
while the bootstrap-estimated median values and the 1--$\sigma$ ranges  
lead to:
\begin{eqnarray}
M_V + 3.33 \, \log P & = & 1.94^{+0.14}_{-0.17} 
     \,(V-I)_0 + 2.72^{+0.18}_{-0.14}  \label{eq9} \\
M_V + 3.33 \, \log P & = & 1.50^{+0.19}_{-0.20} 
     \,(B-V)_0 + 3.41^{+0.16}_{-0.16}
  \label{eq10}
\end{eqnarray}

We suggest that until better and more extensive data become available,
Eqs.~(\ref{eq3})--(\ref{eq4}) or Eqs.~(\ref{eq7})--(\ref{eq8}) 
serve as the preliminary calibrations of the PLC-$\beta$ relation.

\subsection{Limitations of the PLC-$\beta$ relation}
 
One of the limitations of the PLC-$\beta$ relation
is the necessity to limit selection of the systems
to redder than $(V-I)_0=0.25$. There a few reasons why the blue stars
may deviate from the general trends:
\begin{itemize}
\item The underpinning of the 
PLC-$\beta$ relation is the assumption that
the large, Roche-lobe filling component dominates the photometric
characteristics of the system. Only then one can expect 
a tight relation between the size of this component 
(governed by the Roche lobe scaled with the size of the
whole system) and the total luminosity of the
binary system. Even a relatively small, but hot and hence possibly
luminous companion can modify the color index to make it blue.
Thus, blue color index may be considered as 
a direct indication that the hot companion of the 
Roche-filling star is contributing substantially to 
the system light. The well-observed 
system LP\#9 (OGLE-II\#3.134555) which shows clear eclipses
is an excellent illustration of the point in that 
it is one of the blue systems removed from the final
calibration.
\item It is well known that the optical-range
color indices, such as $(B-V)$ or $(V-I)$, lose sensitivity to
effective temperature for very hot stars. The serious problems
in the color -- effective temperature transformations
occur at $\log T_{eff} > 4.5$ (\cite{mas95}). In our case, the blue
stars which deviate from the PLC-$\beta$ relation are not that hot,
and have temperatures within the range $3.9 < \log T_{eff} < 4.0$. 
\item $\beta$~Lyrae is itself a good case
of a blue binary system which does not obey our PLC-$\beta$ relation. 
In this case, it is not the contribution of the other
component to the total light, but apparently the Roche-lobe
filling component is too luminous.
\cite{dob85} summarize the parameters of relevance to our
discussion: $P = 12.937$ days, $M_V = -4.7$ for both components
and $M_V^{vis} = -4.1$ for the visible component, $(B-V)_0=-0.13$
(which implies $(V-I)_0 \simeq -0.1$. Our calibrations
predict $M_V^{VI} = -1.3$ and $M_V^{BV}=-0.5$ for the
visible component. This component is thus
some 2.8 -- 3.6 mag.\ over-luminous relative to the calibrations. 
\item Possibly, use of single
linear relations for the color dependence is simply inappropriate over
the whole wide range $0 < (V-I)_0 < 1.6$ observed for the 
LP $\beta$~Lyrae systems. The five hot systems in our sample do
show the steep relation between $M_V$ and $\log P$. If we lump
all five and use their average color indices 
($(V-I)_0=+0.109$ and $(B-V)_0=-0.067$) to enter these
in Eqs.~(\ref{eq3}) and (\ref{eq4}), then $M_V \propto -3.10 \, \log P$,
with the zero points,
 $+0.87$ and $+1.19$ for the $(V-I)$ and $(B-V)$ based relations,
respectively. These zero points are very different than for the
more common yellow-red LP systems, leading to
relations by about 2 to 2.5 magnitudes
brighter. This would go some way towards expaining the 
over-luminosity of the $\beta$~Lyrae less-massive component. 
\end{itemize}

A serious limitation of any approach requiring de-reddened colour
indices is the uncertainty of 
the reddening correction. Although the colour-index
terms have relatively small coefficients and the strongest
dependence is in the $\log P$ term, nevertheless the reddening
corrections must be applied to the observed data. 
For the SMC OGLE-II data, two indices are
available, $(B-V)$ and $(V-I)$. Can one use form a reddening-free
index, similar to the well-known $Q$ index? Unfortunately, this
is not the case for the particular combination of the indices.
An index constructed in the form: $Q' = (B-V) - k \,(V-I)$, where
$k = E_{B-V}/E_{V-I}$ is not a monotonic function of the indices:
It shows a shallow increase from $(V-I)_0=-0.2$ to 0.9, 
then a broad maximum around $0.9 < (V-I)_0 < 1.4$, and a steep 
decline further to the red. Thus, while the conventional
$Q$-index technique based on the $(U-B)$ and $(B-V)$ would work, 
the one based on the $(B-V)$ and $(V-I)$ indices would fail.
The reddening correction uncertainty will remain the main limitation
of the PLC-$\beta$ relation.

\subsection{External checks: The LMC and M31 data}

Undoubtedly, calibrations of the PLC-$\beta$
relation will be substantially 
improved when similar OGLE-II data for the Large Magellanic 
Cloud become available. At present, support
from other sources is weak. As has been pointed before, 
the currently available MACHO data for the LP
systems in LMC (\cite{macho}) are in a 
non-standard photometric systems only approximating the 
$VR$ filter system. Seven systems in this sample 
have orbital periods longer than 10 days, in the range $12<P<52$ 
days. Unfortunately, all seven are quite blue
($V-R <0.35$, see their location in right panel of Figure~\ref{fig2}).
It is nevertheless possible to establish 
a tight linear relation, similar to the 
calibration Equations (\ref{eq3}) -- (\ref{eq10}), for these systems.
The relation ($M_V = -0.26 \, \log P 
+ 4.31 \, (V-R)_0 -1.58$ for $m-M=18.3$ and $E_{V-R}=0.2$)
has a small dispersion $\sigma = 0.1$ mag., but 
we do not believe that this relation is of much
validity because the photometric system is non-standard 
and all systems are blue.

The data for binary systems in the Andromeda galaxy detected by
the DIRECT project also do not provide an external check
on the calibration of the PLC-$\beta$ relation. 
The project found four LP systems of $\beta$~Lyrae-type in M31
(\cite{DirA}, \cite{DirC}, \cite{DirD}). All four have 
periods within 11 to 14 days, so that they  
do not probe the whole span of the orbital periods 
of the SMC sample. The DIRECT
systems are: D31A--6423, D31C--13944, D31D--5186, 
D31D--7221 (we will call them below by 
consecutive numbers, 1 -- 4). 
The systems \#1 and \#4 
have very poor photometric data (especially 
the crucial color indices are poorly 
determined) while the system \#2 has too 
blue color index (uncorrected $V-I=0.13$) to 
belong to our proper LP sequence. Only the 
third system, D31D-5186 would qualify, with $P=11.80$ days,
$V_{max}=19.5$, $V-I=0.41$ and the very uncertain 
$B-V=-0.2$. The directly evaluated value of $M_V$, 
assuming $m-M=24.5$ (\cite{sta98})
and $E_{B-V}=0.1$ is $M_V=-5.3$, 
while the one estimated from Eq.~(\ref{eq3}), assuming 
the same reddening, is $M_V=-0.2$. Thus, 
the discrepancy between the two estimates is very large, 
either shedding doubt on our SMC calibration or
indicating that the companion is over-luminous (as in the
very case of $\beta$~Lyrae).  
We note that the DIRECT project can discover only 
the very brightest binaries in M31 and these would be 
the ones which would be most likely the
ones which deviate the most from the 
calibration. Also, the color used is highly dependent on
the assumed reddening which is only guessed in this case. 
We cannot exclude a possibility that $(V-I)_0$ 
for the system \#3 is smaller than our threshold of 0.25; 
then the system would belong to the blue group which 
we explicitly excluded from consideration when 
establishing the PLC-$\beta$ relation. 

\section{LONG-PERIOD SYSTEMS IN BAADE'S WINDOW}
\label{bw}

It has been realized recently that
the OGLE-I photometry for Baade's Window (BW)
suffered from systematic zero point offsets 
and linearity problems resulting from the
non-linearity of the original CCD detector. 
At the brightness level of the 
Bulge Red Clump Giants, the offsets are moderate: 
0.021 in $V$, 0.035 in $I$ and thus 0.056 
in $(V-I)$ (\cite{bp99}). More worrisome are 
however the non-linearities, relatively larger in 
the $V$-filter data for red stars and affecting 
particularly strongly the $(V-I)$ color indices of 
faint and red stars. Dr.\ Udalski (private 
communication) estimated that deviations in $(V-I)$ 
may be as large as 0.2 mag.\ for faintest surveyed
stars. Since the color indices play an important
role in our absolute-magnitude calibrations, the
deficiency of the OGLE-I photometry will require
a more extensive re-discussion of the results
of Papers~II and III.

The OGLE project, currently in its Phase~II, is
characterizing the systematic errors of OGLE-I. Some stars
already have excellent new data in $V$, $(B-V)$ and
$(V-I)$. At our request, Dr.\ Udalski kindly checked
the availability of new data for the sequence of the
systems with $P > 1$ days which had been discussed in
\cite{ogle3} (Paper~III). About two thirds of these stars have new
photometry. Among the four BW systems with $P>10$ days,
three have new photometric data. These have been made available
to us in terms of the average light levels, 
requiring additional corrections to the maximum light level. 
They are based on typically 30 observations in $I$ and 
10 observations in $V$. We applied the differences 
$\Delta I = \overline{I_1} - I_1$,
as found on the basis of the OGLE-I light curves 
(Paper~III), to convert the data
from the average ($\overline{I_2}$) to the
maximum ($I_2$) light levels. The details are given 
in Table~\ref{tab2}. 

The left panel of Figure~\ref{fig2}
shows the OGLE-II data for the three LP $\beta$~Lyrae systems
in the period-color plane, with the
assumed reddening corrections the same as in Paper~III
(the three systems are marked by filled asterisks). 
The corrections were found on the basis of an 
assumption that the reddened stars are further than 2 kpc and thus 
suffer the same amount of reddening as the 
Bulge stars in the BW (\cite{sta96}). This assumption can be checked
for consistency using the calibration in Eq.~(\ref{eq3}).   
The estimates of the distances are  
given in the last column of Table~\ref{tab2}. The errors 
reflect the uncertainty of the calibration equation
of about 0.3 mag. The results are consistent
with two systems in the Bulge (possibly on the other
side of it) and one system possibly in front of the Bulge
or inside its elongated part directed toward us. 
The fourth system, \#0.036, without the new data, also appears
to be located in the Bulge. We treat these distance determinations
as very preliminary and disregard a possible metallicity
mismatch between the calibration established for the SMC
systems and the possibly metal-rich systems in Baade's
Window. 

\section{SPATIAL FREQUENCY}
\label{freq}

The spatial frequency of the long-period $\beta$~Lyrae-type 
systems is difficult to estimate because it is practically impossible
to estimate the OGLE-II search volume in the SMC.
In terms of the intrinsic luminosities, the LP $\beta$~Lyrae-type 
systems occupy the range $-5 < M_V < +1$ 
(the calibrated part extends to $M_V \simeq -3$) so that they can be
compared with the Cepheid variables. The OGLE-II project
observed 2155 Cepheids in the SMC
(Dr.\ A.\ Udalski -- private information) which can be compared
with 38 long-period $\beta$~Lyrae-type systems discussed here. 
While a large fraction were
previous discoveries and the LP binary systems are new, this
gives a rough idea about the frequency of occurrence of the
LP $\beta$~Lyrae-type more than 50 times below the spatial frequency
of the Cepheids. As an aside, notice when comparing 
the two types of variables,
that the period term in the LPC-$\beta$ relation 
appears to be close to the expected $M_V \propto -3.33 \, \log P$,
while the Cepheids show a period dependence,
with the period coefficient $-2.78$ (\cite{fw87}) or $-2.76$
(\cite{mf91}). The similarity of these dependencies should not be
surprising as both are driven by the same dimension -- response time
relationship of the dynamical time scale. 
 
The Baade's Window data
give us only a very approximate constraint on their frequency in our
Galaxy. If one determines the average frequency of occurrence by simply
dividing their number (four) by the volume of the OGLE-I search
in the BW area (about $2.5 \times 10^7$ pc$^3$ to the Galactic
center), then an estimate of the frequency is
$1.6 \times 10^{-7}$ LP systems per cubic parsec. However,
the LP $\beta$~Lyrae systems appear to be
strongly concentrated towards the Galactic Bulge so that
such a naive estimate does not take into account the increase
in numbers of stars as the line of sight penetrates the Galactic
Bulge. We recall that the W~UMa-type systems of the disk
population approximately follow the stellar density in the BW
direction with the average density of about $7.6 \times 10^{-5}$
systems per cubic parsec (\cite{open}).

\section{DISCUSSION}
\label{disc}

This paper discusses consequence and observational 
support for an assumption that the long-period ($P > 10$ days), 
apparently contact binary systems observed in Baade's Window, 
and in LMC and SMC, are in fact $\beta$~Lyrae systems with 
the cool components dominating their total luminosity. 
We analyzed in detail the LP systems in the SMC 
and found that they form a well-defined, moderately tight, 
``inverted'' period-color relation, implying a color--size 
relation characteristic for giants rather than for 
main-sequence stars. The principal support,
however, comes from the existence of 
a period-color-luminosity relation for those which are 
sufficiently cool. Tentatively, we set the color-index limit at 
$(V-I)_0>0.25$. An absolute-magnitude calibration of the
new PLC-$\beta$ relation in the 
form $M_V=M_V(\log P, \, color)$, established 
in a preliminary form on the basis of 33 systems in the SMC, 
has the expected slope $M_V \propto -3.33 \, \log P$. It
offers an exciting application for distance determination
to nearby galaxies. The LP $\beta$~Lyrae systems are the 
brightest among binary stars and their continuous light variations 
make them also very easy to spot. They occupy roughly
the same range of the absolute magnitudes as the Cepheid variables,
but appear to be by about 50 times less common in space.
Therefore, the PLC-$\beta$ relation can play only an auxiliary --
although entirely independent -- role in distance determinations.

We see sequences of ostensibly 
contact, long-period systems not only in the SMC, but
also in Baade's Window and in MACHO data for LMC.
However, the latter data sets could not be used 
to improve the absolute magnitude 
calibration of the PLC-$\beta$ relation 
for reasons described in detail in Sections
\ref{pc} and \ref{bw}.
It is expected that the data for LMC currently collected 
by the OGLE-II project will provide an excellent check and 
improvement of the tentative results presented here.

We conclude by observing that the $\beta$~Lyrae binaries
discussed in this paper are expected to show some
similarities to the cataclysmic variable (CV) binaries.
The CV binaries are also known to have secondary (less massive
components) filling their Roche lobes. As was observed
by \cite{faul72}, a combination of Kepler's law,
$a^3 \propto (M_1+M_2)\,P^2$, with the
approximate relation for the relative radius of the Roche
lobe filling star, $R_2/a = 0.462 (q/(1+q))^{1/3}$, 
with $q = M_2/M_1 \le 1$, gives a relation similar to that
for pulsating stars: $P = 0.438 /\sqrt{\rho}$, where the period
$P$ is in days and the mean density 
of the star, $\rho$, is in g~cm$^{-3}$.
This relation is valid for small mass ratios ($q < 0.5$),
but corrections for larger values of $q$ are small. Because
the range of the observed period for the LP sequence in the
SMC is large, one expects the Roche-lobe
filling components to have mean
densities within a large range from
about $2 \times 10^{-3}$ g~cm$^{-3}$ to
less than $10^{-5}$ g~cm$^{-3}$, that is within typical
range observed for yellow-red giants and supergiants.

\acknowledgements
Special thanks are due to Dr.\ Andrzej Udalski for his prompt
response -- in spite of many ongoing activities --
to a request to extract new OGLE-II data for the
LP systems observed by OGLE-I project in Baade's Window and
for several explanatory comments on the data.
Thanks are also due to Dr.\ Stefan Mochnacki for very useful
comments and suggestions.
The authors would like to thank the OGLE team for making the
data freely available over the computer networks.

CM acknowledges the support of the Italian Ministry of Research
(MURST) through a Cofin98 grant and SMR acknowledges the support
of the Natural Sciences and Engineering Research 
Council (NSERC) of Canada.

\newpage

\noindent
Captions to figures:

\figcaption[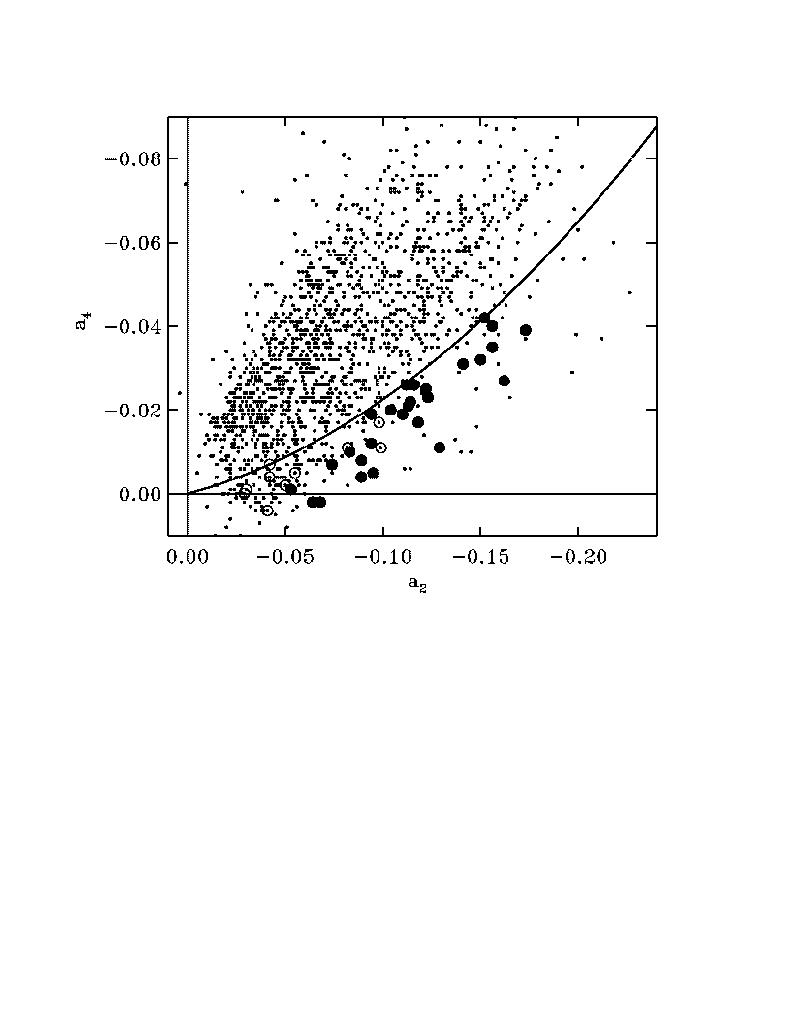]{\label{fig1}
The Fourier light-curve shape filter selects contact
binaries by comparison of the $a_2$ and $a_4$
cosine coefficients. Systems with properties of
contact binaries are located below the curved 
line. The small dots give the data for all eclipsing
systems in the SMC while the large filled circles mark 38
long-period ($P > 10$ days) 
contact systems discussed in this paper. The open circles
are used for 11 systems with relatively poorer light curves
which have been given half weight in luminosity
calibrations discussed further in the paper.
Note that contrary to the results
for Baade's Window (\cite{ogle1}), the contact
domain is rather thinly populated in the SMC, mostly due to
the absence of the genuine W~UMa-type binaries ($P<1$ day). 
We give arguments in the paper that the apparently contact 
long-period systems are in fact semi-detached
binaries of the $\beta$~Lyrae-type with component-distortion
effects dominating over shallow or absent eclipses. 
}

\figcaption[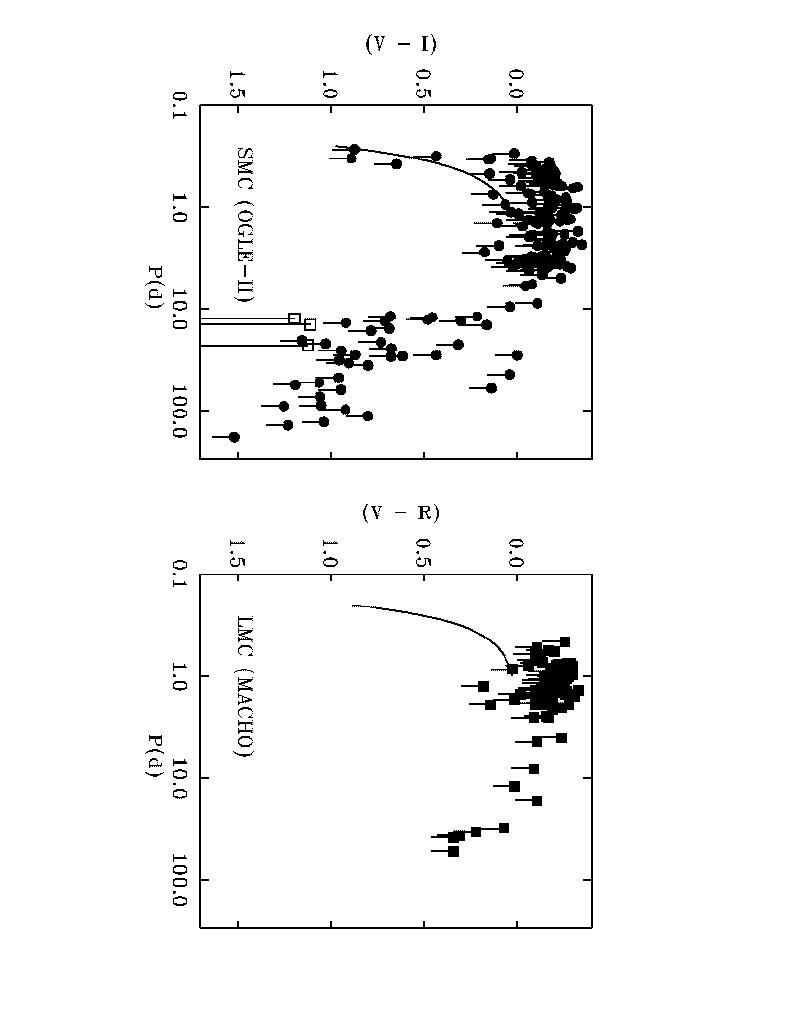]{\label{fig2}
The period--color diagram for all contact systems in the
SMC (this paper, left panel) and LMC (the MACHO data, right
panel, as discussed in Rucinski 1999). The large circles
show systems discussed in this paper; the open circles mark those
which have been assigned half weights in luminosity calibrations
discussed later on. The SMC data are based on the $(V-I)$ 
color index while the LMC data are based on the
$(V-R)$ color index. The curves give the blue short-period 
envelope for normal, short-period 
W~UMa-type systems. For simplicity, we have assumed the same 
reddening vectors for all systems in the SMC, 
$E_{V-I}=0.12$ (vertical lines). In subsequent sections
we find that systems with $(V-I)0>0.25$ provide a 
well defined period -- luminosity -- color relation. 
Three large asterisks show locations of the long-period
systems in Baade's Window; they are more fully discussed in
Section~\ref{bw}. We cannot directly utilize the
BW or LMC data in this paper, but point out the existence 
of the LP systems which appear to be in 
contact and form a sequence extending to long orbital 
periods. 
}

\figcaption[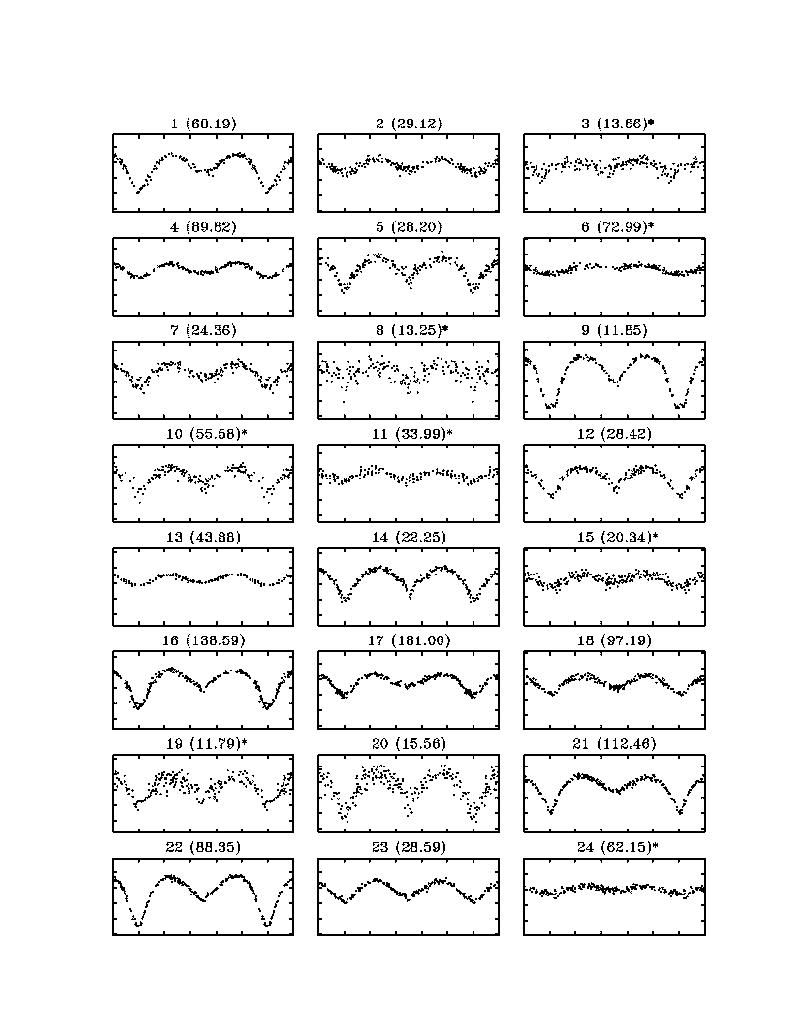]{\label{fig3}
A collection of $I$ band light curves for 38 long period 
contact systems in the SMC. This figure contains 24 light curves;
the remaining curves are in the next figure. The systems are identified
by numbers above each of the panels with the orbital
periods given in parentheses. An asterisk added to the
star number indicates half-weight given in the luminosity
calibrations. The vertical scale
of each panel spans one magnitude and extends from 0.6 mag.\
below to 0.4 mag.\ above the average magnitude. The maximum light
magnitudes, the orbital periods, the color data as well
as the Fourier cosine coefficients are given in Table~1.}

\figcaption[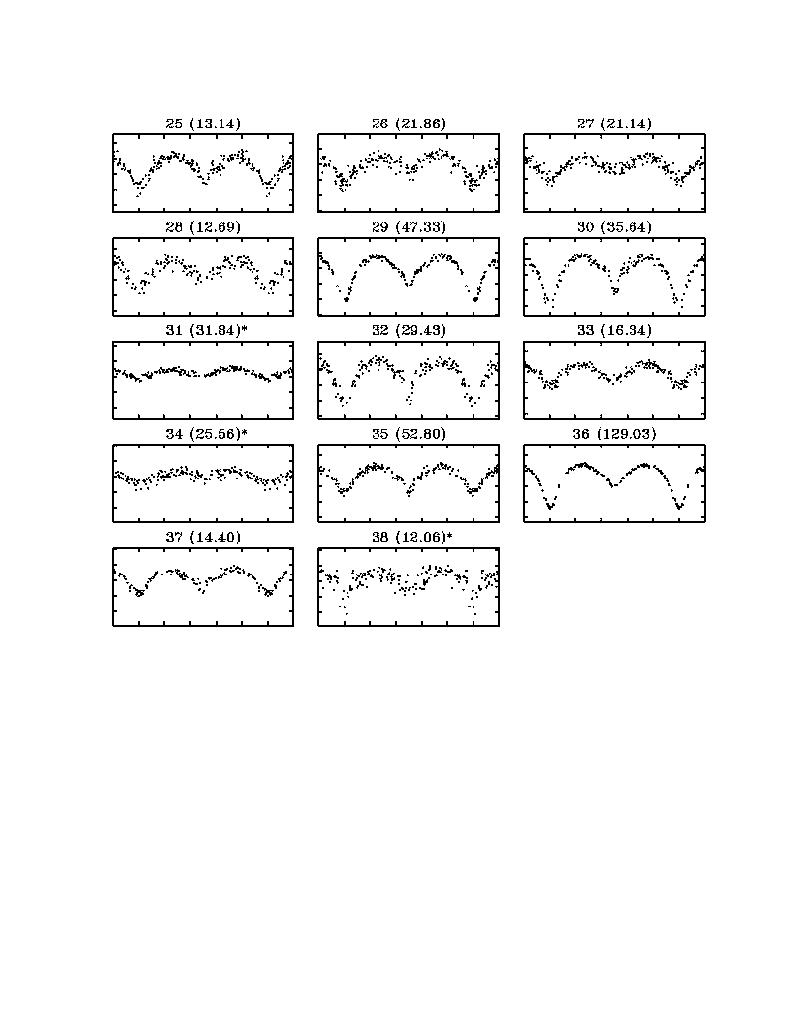]{\label{fig4}
The same as for Figure~3, but for 14 among 38 long period 
contact systems in the SMC.}

\figcaption[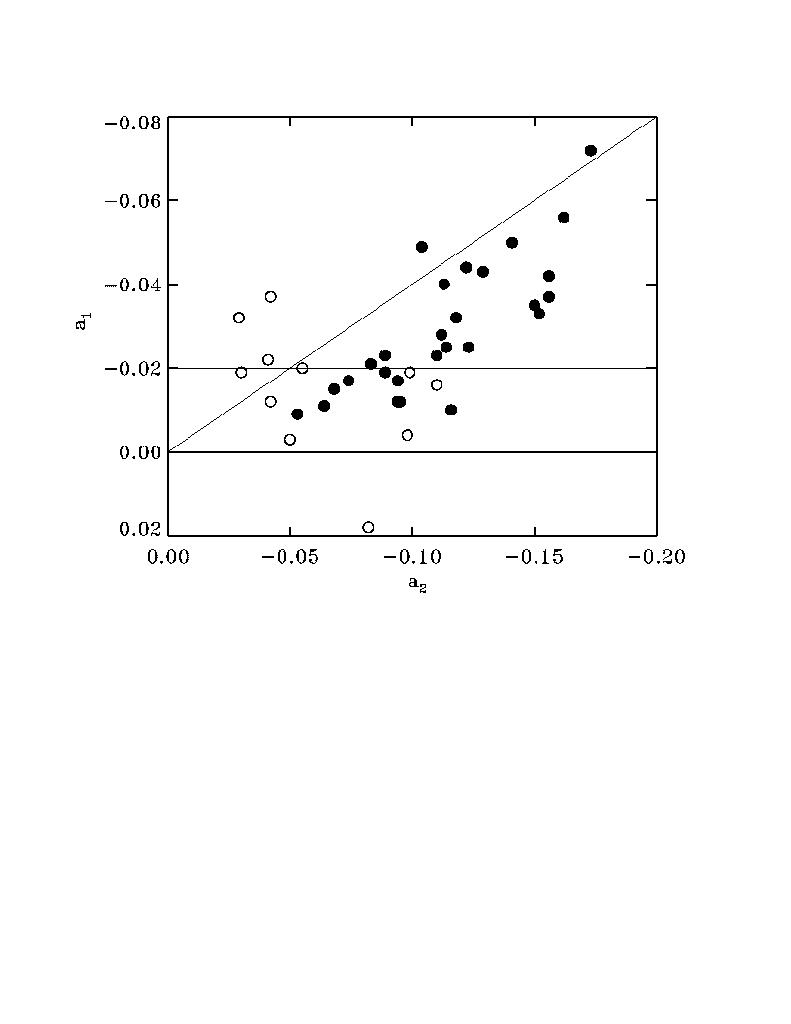]{\label{fig5}
The first cosine term of the light-curve Fourier decomposition, 
$a_1$, appears to scale in proportion to the largest 
term, $a_2$. This confirms that the light curves 
are very similar in shape. The open circles mark systems with
half weights in the calibration solutions.
}

\figcaption[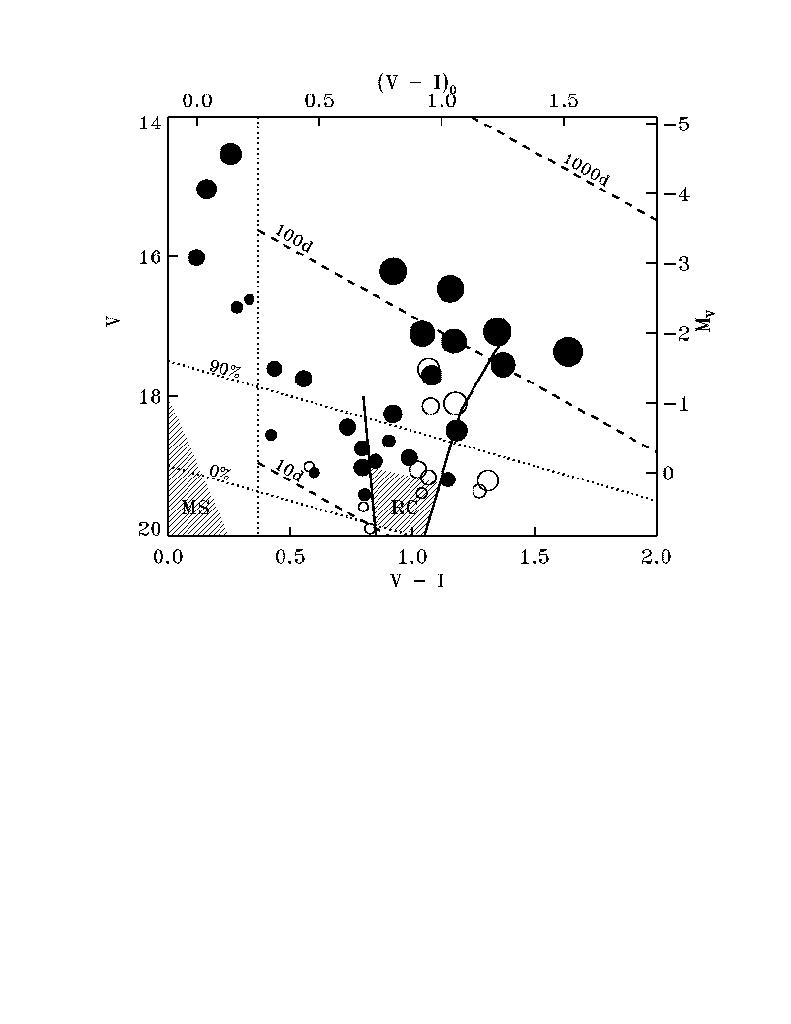]{\label{fig6}
The color-magnitude diagram for the LP systems in SMC.
The orbital periods of individual systems 
are marked by the size of the circle.
The right and upper axes show the de-reddened color index
and absolute magnitude data for
$E_{V-I} = 0.12$ and $(m-M)_0 = 18.8$. 
The slanted dotted lines give the approximate locations of the 
90\% and 0\% completeness levels of the OGLE-II variable star 
search. The broken lines represent the constant period lines
according to the calibration of our PLC-$\beta$ relation as
in Eq.~(7). They extend to the red of $(V-I)_0=0.25$. The
current calibrations are not valid to the blue of that point
(marked by the vertical dotted line),
although a period dependence with a correct slope 
is observed for such systems as well. 
The main sequence (MS) and the red clump (RC) of the SMC 
are marked by shading.}

\figcaption[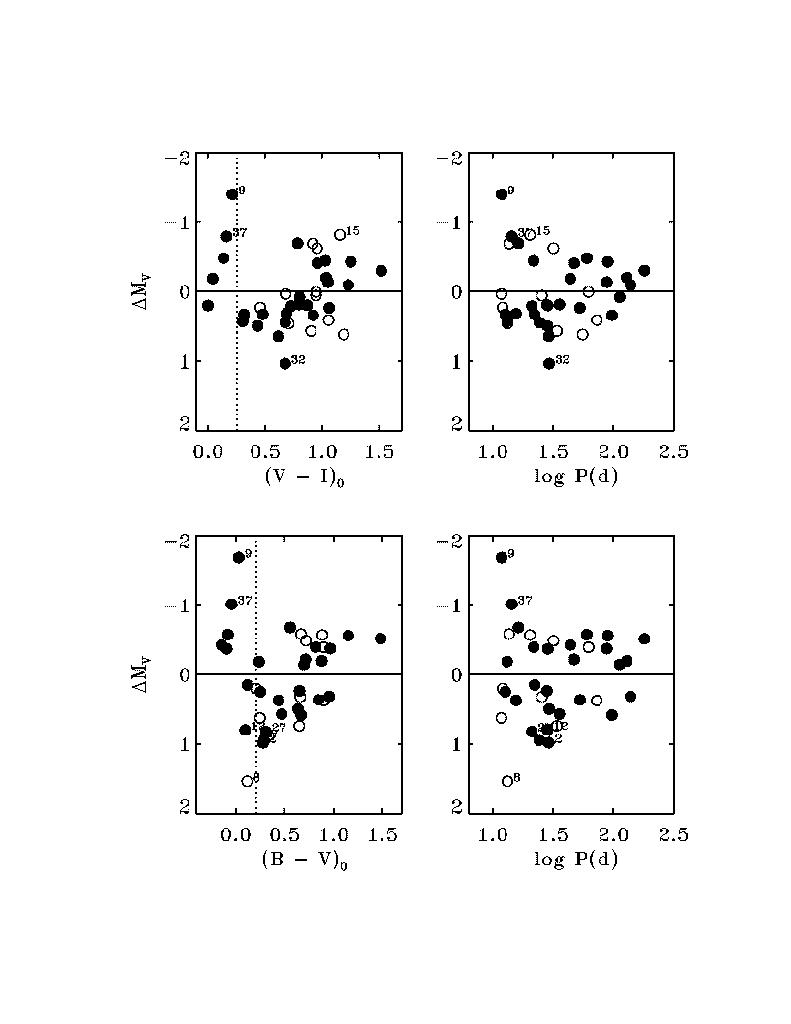]{\label{fig7}
The deviations from the regression lines of the 
calibration in Equations (1) -- (2).
Note the large negative deviations at blue color indices.
The labels are the consecutive numbers in Table~1; they are
given for systems deviating more than 0.75 mag.\ from
the linear regression. Note in particular the location of the
system LP\#9 which is discussed in the text. 
The dotted lines give the color-index cutoff lines at
$(V-I)_0=0.25$ and the corresponding $(B-V)_0=0.21$. 
}

\figcaption[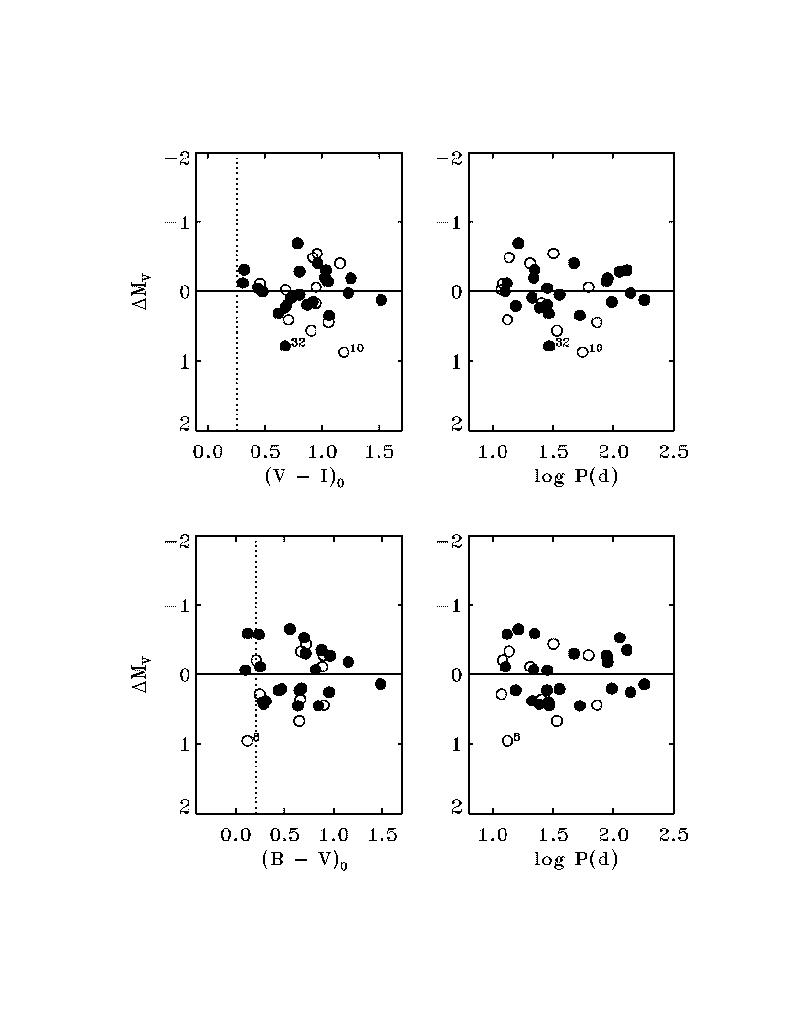]{\label{fig8}
Same as in Figure~7, but with a limitation
to systems with $(V-I)0>0.25$, as in Equations (3) -- (4). 
Notice a small inconsistency -- which we disregarded in our
determinations of the PLC calibrations -- 
in that 3 systems are slightly bluer
than $(B-V)_0=0.21$ which roughly corresponds to 
the cutoff at $(V-I)_0=0.25$.
}

\clearpage

\begin{table}                 
\dummytable \label{tab1}      
\end{table}

\begin{table}                 %
\dummytable \label{tab2}      
\end{table}


\begin{thebibliography}{}

\bibitem[De Greve \& Linnell 1994]{deG94}
   De Greve, J.P., \& Linnell, A.P. 
   1994, \aap, 291, 786
\bibitem[Dobias \& Plavec 1985]{dob85}
   Dobias, J.J. \& Plavec, M.J. 1985, \aj, 90, 773
\bibitem[Faulkner et al.\ 1972]{faul72}
   Faulkner, J., Flannery, B.P., \& Warner, B.
   1972, \apj, 175, L79
\bibitem[Feast \& Walker 1987]{fw87}
   Feast, M.W. \& Walker, A.R. 1987, \araa, 25, 345
\bibitem[Flannery 1976]{fla76}
   Flannery B.P., 1976, ApJ, 205, 217
\bibitem[Gibson 1999]{gib99}
   Gibson, B.K. 
   1999, Mem.Soc.Astr.Ital., in press; astro-ph/9910574
\bibitem[Harmanec \& Scholtz 1993]{har93}
   Harmanec, P., \& Scholtz, G. 
   1993, \aap, 279, 131
\bibitem[Huang 1963]{hua63}
   Hunag, S.--S. 1963, \apj, 138, 342
\bibitem[Hubeny \& Plavec 1991]{hub91}
   Hubeny, I., \& Plavec, M.J. 
   1991, \aj, 102, 1156
\bibitem[Kaluzny et al.\ 1999]{DirD}
   Kaluzny, J., Mochejska, B.J., Stanek, K.Z., 
   Krockenberger, M., Sasselov, D.D.
   Tonry, J.L., \& Mateo, M. 
   1999, \apj, 118, 346
\bibitem[Lucy 1976]{luc76}
   Lucy L. B., 1976, ApJ, 205, 208
\bibitem[Lucy \& Wilson 1979]{LW79}
   Lucy L. B., Wilson R. E., 1979, ApJ, 231, 502
\bibitem[Maceroni \& Rucinski 1999]{mr99} 
   Maceroni, C., \& Rucinski, S.M. 
   1999, \aj, 118, 1819 (Paper IV)
\bibitem[Madore \& Friedman 1991]{mf91}
   Madore, B.F. \& Friedman, W.L. 1991, \pasp, 103, 667
\bibitem[Massey et al.\ 1995]{mas95}
   Massey, P., Lang, C.C., DeGioia-Eastwood, K., \&
   Garmany, C.D. 1995, \apj, 438, 188
\bibitem[Nelson et al.\ 2000]{nel2K}
   Nelson, C.A., Cook, K.H., Popowski, P., \& Alves, D.R.
   2000, \aj, 119, 1205
\bibitem[Paczynski et al.\ 1999]{bp99}
   Paczynski, B., Udalski, A., Szymanski, M., Kubiak, M., 
   Pietrzynski, G.,Soszynski, I., Wozniak, \& P. Zebrun, K. 
   1999, Acta Astron., 49, 319       
\bibitem[Plavec 1985]{pla85}
   Plavec, M.J. 1985, in Interacting Binaries, edited by
   P.P.\ Eggleton and J.E.\ Pringle (Reidel, Dordrecht), p.~155
\bibitem[Popowski 2000]{pop2K}
   Popowski, P. 2000, \apj, 528, L9
\bibitem[Robertson \& Eggleton 1977]{RE77}
   Robertson J.A., Eggleton P.P., 1976, MNRAS, 179, 359
\bibitem[Rucinski 1994]{cal1} 
   Rucinski, S.M. 
   1994, \pasp, 106, 462         
\bibitem[Rucinski 1995]{cal2}
   Rucinski, S.M. 
   1995, \pasp, 107, 648         
\bibitem[Rucinski 1997a]{ogle1}
   Rucinski, S.M. 
   1997a, \aj, 113, 407 (Paper I)         
\bibitem[Rucinski 1997b]{ogle2}
   Rucinski, S.M. 
   1997b, \aj, 113, 1112 (Paper II)        
\bibitem[Rucinski 1998a]{ogle3}
   Rucinski, S.M. 
   1998a, \aj, 115, 1135 (Paper III)        
\bibitem[Rucinski 1998b]{open}
   Rucinski, S.M. 
   1998b, \aj, 116, 2998                   
\bibitem[Rucinski 1999]{macho}
   Rucinski, S.M. 
   1999, Acta Astr., 49, 341     
\bibitem[Rucinski \& Duerbeck 1997]{rd97}
   Rucinski, S.M., \& Duerbeck, H.W. 
   1997, \pasp, 109, 1340
\bibitem[Stanek 1996]{sta96}
   Stanek, K.Z. 
   1996, \apj, 460, L37
\bibitem[Stanek \& Garnavich 1998]{sta98}
   Stanek, K.Z., \& Garnavich, P.M. 
   1998, \apj, 503, L131
\bibitem[Stanek et al.\ 1998]{DirA}
   Stanek, K.Z., Kaluzny, J., Krockenberger, M., Sasselov, D.D.
   Tonry, J.L., \& Mateo, M. 
   1998, \apj, 115, 1894
\bibitem[Stanek et al.\ 1999]{DirC}
   Stanek, K.Z., Kaluzny, J., Krockenberger, M., Sasselov, D.D.
   Tonry, J.L., \& Mateo, M. 
   1999, \apj, 117, 2810
\bibitem[Udalski 1998]{uda98}
   Udalski, A. 
   1998, Acta Astron., 48, 113
\bibitem[Udalski et al.\ 1997]{ogleII}
   Udalski, A., Kubiak, M., \& Szymanski, M. 
   1997, Acta Astron., 47, 319            
\bibitem[Udalski et al.\ 1998a]{uda98a}
   Udalski, A., Soszynski, I., Szymanksi, M., Kubiak, M.,
   Pietrzynski, G., Wozniak, P., \& Zebrun, K. 
   1998a, Acta Astron., 48, 563             
\bibitem[Udalski et al.\ 1998b]{uda98b}
   Udalski, A., Pietrzynski, G., Wozniak, P., Szymanski, M.,
   Kubiak, M., \& Zebrun, K. 1998b, \apj, 509, L25  
\bibitem[Wilson 1974]{wil74}
   Wilson, R.E. 1974, \apj, 189, 319
\bibitem[Woolf 1965]{woo65}
   Woolf, N.J. 1965, \apj, 141, 155

\end{thebibliography}
\end{document}


\pagestyle{empty}

\begin{deluxetable}{ccccccccccccc}
\small
\tablecolumns{13}
\tablewidth{0pt}
\tablenum{2}
\tablecaption{OGLE-II Data for $\beta$~Lyrae systems with $P > 10$ days
in Baade's Window}
\tablehead{
  \colhead{OGLE-I} &          
  \colhead{OGLE-II}  &        
  \colhead{P(d)}  &             
  \colhead{$E_{V-I}$}  &        
  \colhead{$\overline{I}_2$} &         
  \colhead{$\overline{(V-I)}_2$} &         
  \colhead{$\Delta I_1$} &      
  \colhead{$I_1$} &             
  \colhead{$(V-I)_1$} &         
  \colhead{$I_2$} &             
  \colhead{$V_2$} &             
  \colhead{$M_V^{cal}$} &       
  \colhead{d(kpc)}           
}                  
\startdata
0.036 & \nodata   & 26.307 & 0.70 &\nodata & \nodata & \nodata & 15.51 & 1.82 & \nodata &(17.33)& (0.21) & ($11.8 \pm 1.8$) \nl
3.051 & 45.501074 & 14.243 & 0.77 & 16.08 & 1.88 & 0.05 & 15.98 & 1.83 & 16.03 & 17.91 & 1.11 & $9.4 \pm 1.2$ \nl
3.079 & 46.036352 & 12.313 & 0.87 & 16.51 & 2.06 & 0.05 & 16.43 & 2.07 & 16.46 & 18.52 & 1.49 & $9.3 \pm 1.2$ \nl
4.002 & 46.135504 & 22.672 & 0.67 & 14.59 & 1.79 & 0.18 & 14.33 & 1.99 & 14.41 & 16.20 & 0.44 & $6.6 \pm 0.8$ \nl
\tablecomments{The first two columns give the star identifications
consisting of the respecive OGLE field numbers  
(integer part) and of the number in the field (fractional part).
The reddening estimate for $d>2 kpc$, 
$E_{V-I}$, and the maximum light values of
$I_1$ and $(V-I)_1$ are from Paper~III. The
differences between the mean and maximum levels, $\Delta I_1$, have
been calculated on the basis of the OGLE-I light curves. The
values of $\Delta I_1$ have
been used to convert the average OGLE-II data to maximum light level
values, $I_2$. It has been assumed that the color indices
$(V-I)_2$ are the same for the average and maximum light levels.
The values of $M_V^{cal}$ have been calculated using Eq.~(3) leading
to the distance estimates through: 
$V-M_V^{cal}=5 \log d - 5 + 2.5\,E_{V-I}$.
The OGLE-II data are not available for the system \#0.036; in this case 
the distance estimate is based on the OGLE-I data. 
}
\enddata
\end{deluxetable}